\documentclass[copyright,creativecommons]{eptcs}
\providecommand{\event}{TFPIE 2023}
\usepackage[utf8]{inputenc}
\usepackage[T1]{fontenc}
\usepackage{underscore}
\usepackage{graphicx}
\usepackage{microtype}
\usepackage{cite}
\usepackage{xspace}
\graphicspath{ {./images/} }
\usepackage{tikz}
\usetikzlibrary{positioning,fit,arrows.meta,backgrounds}
\tikzset{
    module/.style={%
        draw, rounded corners,
        minimum width=#1,
        minimum height=7mm,
        font=\sffamily
        },
    module/.default=2cm
}

\newcommand{\ProofBuddy}{\textsc{ProofBuddy}\xspace}

\title{\texorpdfstring{\ProofBuddy: \\A Proof Assistant for Learning and Monitoring}{ProofBuddy: A Proof Assistant for Learning and Monitoring}}

\author{Nadine Karsten\institute{Technische Universität Berlin \\ Berlin, Germany}\email{n.karsten@tu-berlin.de}\and Frederik Krogsdal Jacobsen\institute{Technical University of Denmark \\ Kongens Lyngby, Denmark}\email{fkjac@dtu.dk} \and Kim Jana Eiken\institute{Technische Universität Berlin \\ Berlin, Germany}\email{k.eiken@campus.tu-berlin.de} \and Uwe Nestmann\institute{Technische Universität Berlin \\ Berlin, Germany}\email{uwe.nestmann@tu-berlin.de} \and Jørgen Villadsen\institute{Technical University of Denmark \\ Kongens Lyngby, Denmark}\email{jovi@dtu.dk}}

\def\titlerunning{\ProofBuddy: A Proof Assistant for Learning and Monitoring}
\def\authorrunning{N. Karsten, F. K. Jacobsen, K. J. Eiken, U. Nestmann \& J. Villadsen}
\def\copyrightholders{N. Karsten, F. K. Jacobsen, K. J. Eiken,\\ U. Nestmann \& J. Villadsen}

\begin{document}

\maketitle

\begin{abstract}
Proof competence, i.e.~the ability to write and check (mathematical) proofs, is an important skill in Computer Science, but for many students it represents a difficult challenge.
The main issues are the correct use of formal language and the ascertainment of whether proofs, especially the students' own, are complete and correct.
Many authors have suggested using proof assistants to assist in teaching proof competence, but the efficacy of the approach is unclear.
To improve the state of affairs, we introduce \ProofBuddy: a web-based tool using the Isabelle proof assistant which enables researchers to conduct studies of the efficacy of approaches to using proof assistants in education by collecting fine-grained data about the way students interact with proof assistants.
We have performed a preliminary usability study of \ProofBuddy at the Technical University of Denmark.
\end{abstract}

\section{Introduction}
The curriculum of a Bachelor study programme in Computer Science typically includes programming skills, technical understanding about computers, theoretical knowledge and math.
Especially the latter two cause difficulties for students.
Theoretical courses cover formal languages, automata, logic, complexity and computability, which all have in common that they build on mathematical structures and proofs about them.
Hence Computer Science students have to write and verify proofs in several theoretical courses.
This requires \emph{proof competence}, which includes the following four specific competencies \cite{Brunner2014}:
(1) \emph{professional} competence describes the contextual knowledge about the proposition that has to be proved;
(2) \emph{representation} competence is the ability to write down a proof in sufficiently formal language;
(3) \emph{communication} competence is then understood as arguing about the procedure and solution of a proof;
(4) finally, \emph{methodological} competence summarizes three aspects \cite{Heinze2003}: proof scheme, proof structure and chain of conclusions.

All of these competencies are taught as part of introductory courses in Theoretical Computer Science, but proof competence with the above-mentioned facets is usually not an explicit part of the curriculum. 
As a result, students work through proofs that the teacher presents during lectures and try to replicate \cite{SeldenSelden2009}.
Most students fail this way.
In \cite{FredeKnobelsdorf2018}, Frede and Knobelsdorf analyzed the homework and the written exams of an introductory course at Technische Universität Berlin (TUB).
The result was that students make the most mistakes in exercises with proofs, regardless of the final grade.
The main challenges in writing proofs are the usage of formal language while writing proofs and the ascertainment of whether a proof is complete and correct \cite{analysingStudentPracKnobelsdorf16, Kiehn2017, KnobKreitz2017}.
When students try to write proofs in introductory courses in Theoretical Computer Science, the failure rate is therefore high.

As we will see in the next section, proof assistants have often been suggested as a tool to improve proof competence.
A proof assistant can be thought of as a functional programming language combined with a language for reasoning about programs.
This allows users of proof assistants to write proofs in a formal, structured way, and get instant feedback on the correctness of their proofs from the proof checker.
Proponents of using proof assistants in education claim many alleged benefits of the approach, but unfortunately, these claims are rarely supported by more than anecdotal evidence or surveys asking students to report their attitudes about the efficacy of the approach.
Conversely, there are reasonable doubts about the efficacy of the approach: it is not clear that proof competences gained using proof assistants transfer to pen-and-paper proofs, proof assistants may enable a certain flavor of ``brute-force proving'' and introducing students to a proof assistant may use precious teaching time that could be better spent elsewhere.

Clearly, authors should do something to test these claims, but as we will see in the next section this is rarely done, at least partially due to a lack of tools that would enable the studies needed to test them.
Inspired by similar approaches in studying the efficacy of didactic techniques when teaching functional programming and mathematics, we introduce \ProofBuddy: an instrumented version of the Isabelle proof assistant~\cite{Nipkow-Paulson-Wenzel:2002,IsabelleProving}.
The idea of \ProofBuddy is to collect fine-grained data about the interactions of students with the proof assistant and to use this data to conduct studies of the efficacy of various didactic approaches.
Additionally, \ProofBuddy has optional features to attempt to counteract some of the common doubts about the use of proof assistants in education.
We have evaluated the  usability of \ProofBuddy and the data collection capabilities based on a set of research questions that we imagine future didactic studies could want answers to and found that the type of data collected by the tool is sufficient to answer these types of questions.

In summary, our contributions are:
\begin{itemize}
    \item An overview of the current issues in studying the efficacy of using proof assistants in education.
    \item \ProofBuddy: a version of the Isabelle proof assistant with instrumentation for collecting data about user interaction with the proof assistant.
    \item A preliminary evaluation of the usability of \ProofBuddy from the perspective of students using the tool.
    \item A preliminary evaluation of the adequacy of the interaction data collected by \ProofBuddy for conducting didactic studies.
\end{itemize}
In the next section, we will discuss existing uses of proof assistants in education, including the claimed benefits and drawbacks of using proof assistants in education, as well as some of the tools from the domain of functional programming that have inspired the design of \ProofBuddy.
We will outline potential approaches to mitigate the drawbacks and investigate the claimed benefits of using proof assistants in education in \autoref{sec:what-is-to-be-done}.
Next, we describe the features (inspired by these potential approaches) and implementation of \ProofBuddy in \autoref{sec:proofbuddy}.
In \autoref{sec:evaluation} we evaluate \ProofBuddy from two perspectives: the adequacy of the collected data for conducting didactic studies and the usability from the perspective of students.
Finally we outline future work in \autoref{sec:future-work} before concluding in \autoref{sec:conclusion}.

\section{Related Work}\label{sec:rel-work}
This section covers work related to \ProofBuddy, by which we mean not only descriptions of superficially similar tools, but also the articles and studies that have motivated the development of \ProofBuddy, and the tools from similar fields that have inspired its design.
We will cover a selection of reports on using proof assistants in education, then summarize the claimed benefits and observed drawbacks of doing so.
Finally, we will note some approaches from similar fields which have inspired the design of \ProofBuddy.

\subsection{Using Proof Assistants in Education}
The idea to use proof assistants in education is not new, but the scientific results of these efforts have mostly been experience reports, and not rigorous evaluations of the efficacy of the approach.
In this paper we will give only a short overview of approaches.
Several courses target students in higher semesters and assume proof competence and functional programming skills \cite{compuMethaphysics,LSDproofsSemantic}.
Knobelsdorf et al.\ \cite{KnobKreitz2017} report on the use of the proof assistant Coq \cite{bookCoq} to enhance proof competence, with the following result: while students were able to prove theorems with Coq after the course, their ability to write pen-and-paper proofs was not significantly better than before the course.
Böhne and Kreitz \cite{coqToText} therefore attempt to improve pen-and-paper proof competences by requiring students to add comments in the formal Coq proof; nevertheless the improvement in writing pen-and-paper proofs was only minimal.

Some have also tried to integrate a proof assistant in first-year courses.
Avigad \cite{Avigad2019} used the proof assistant Lean~\cite{Lean} in a logic course in 2015 to combine mathematical language and natural language in proofs.
Avigad describes only anecdotal evidence and concludes that it seems necessary to develop quantitative methods and tools for data collection before carrying out a comparative study of the efficacy of the approach.
Thoma and Iannone \cite{Thoma2022} describe a study where students could choose to use Lean in addition to the usual material in a first-year university course.
The small number of students who used Lean in voluntary workshops used more mathematical language during an interview and structured their proofs better.
Because Lean was not mandatory it is possible that only the students who were already doing well in the course chose to also use the proof assistant.
Lean was also used for the Natural Number Game \cite{NaturalNumberGame}, which is a web application designed to teach formal proofs in a proof assistant using basic theorems in arithmetic and logic.
Kreitz, Knobelsdorf and Böhne \cite{inp:BoehneKnobelsdorfKreitz16a} consider a first-semester course where Coq is used in the lectures and the exercises to support the handling of formalisms, but this course was not realized.
The web tool Waterproof by Wemmenhoven et al.\ \cite{waterproof} allows users to write Coq proofs with natural language.
The authors developed a library to teach students in Analysis I where using the proof assistant was voluntary.
The students who used the tool had better grades at the end of the course, but again this may simply be a result of self-selection bias.
At TUB, we integrated Isabelle in a second semester course in 2021.
We used Isabelle to introduce propositional and first-order logic step by step, but using Isabelle for the exercises and homework was optional.
The students had fun in proving but it was time-consuming for them to step into Isabelle.

Lurch was a ``proof-checking word processor'' in which students could write natural language texts and mark certain parts of their input as having mathematical content, which the tool would then check~\cite{Lurch2013a,Lurch2013b}.
The tool focused on good user experience for beginners, and the expressivity of the formalism is unclear.
The desktop version of the tool is no longer available, and the development of a new web-based version seems to have ceased in 2018~\cite{Lurch2017}.
The web tool Carnap \cite{LeachKrouse2017} offers a wide range of exercises in mathematics and logic.
In 2017 there were several courses where different approaches with different students where tested.
The author anecdotally found that Carnap offered exercises appropriate for different levels of competence.
The AProS project is an effort to develop a number of courses and tools to teach logic and proofs~\cite{AProS2007}.
These courses are computer-taught, not only computer-assisted, and include a number of interactive environments which are essentially proof assistants.
Burke provides anecdotal evidence for the efficacy of the AProS project~\cite{Burke2006}.
Clide \cite{dfki6811, dfki7514} is a web application based on Isabelle.
Clide was implemented already in 2013 and consists of an editor integrated within a web interface, and an Isabelle backend for checking proofs.
Clide was not focused on learning, however, but focused on allowing collaborative writing and editing of proofs.
The Clide application seems to no longer be available.

Many purpose-built proof assistants designed for learning specific topics have been developed, including Jape \cite{sufrin1996user, sufrin1998user}, ProofWeb \cite{hendriks2010teaching}, SPA \cite{SPA}, SeCaV \cite{SeCaV} and NaDeA \cite{Villadsen2015,Villadsen2018,Villadsen2019}.

The second and fifth author have previously conducted a study of student experiences using Isabelle, but our methods suffered from a lack of data on actual student behavior and interactions when using the proof assistant~\cite{Jacobsen2022}.
Knobelsdorf et al. limited themselves to manually observing student behavior and categorizing the types of questions asked by students~\cite{KnobKreitz2017}.
Mariotti has conducted a number of studies on learning to prove with a dynamic geometry environment, but similarly describes only manual observations and interviews with students~\cite{Mariotti2019}.

\subsection{Claimed Benefits and Drawbacks of Using Proof Assistants in Education}\label{sec:pas-in-edu}
As described in the previous sections, there have been many attempts to use proof assistants in education.
In this section, we will summarize the claimed benefits and drawbacks of this approach in the literature.

We start with the claimed benefits of using proof assistants in education, of which there are many:
\begin{itemize}
    \item Proof assistants are useful for teaching mathematics~\cite{Avigad2019, inp:BoehneKnobelsdorfKreitz16a, Lurch2013a, Lurch2013b, hendriks2010teaching}
    \item Proof assistants are useful for teaching functional programming~\cite{inp:BoehneKnobelsdorfKreitz16a,coqToText, hendriks2010teaching, Jacobsen2022, KnobKreitz2017, LSDproofsSemantic}
    \item Proof assistants are useful for teaching logic~\cite{Avigad2019,inp:BoehneKnobelsdorfKreitz16a,coqToText, Burke2006, SeCaV, hendriks2010teaching, Jacobsen2022, KnobKreitz2017, LSDproofsSemantic, AProS2007, compuMethaphysics, Villadsen2019}
    \item Proof assistants are useful for teaching abstract thinking~\cite{inp:BoehneKnobelsdorfKreitz16a, Lurch2013b, AProS2007, compuMethaphysics}
    \item Proof assistants make the rules of formal reasoning clear~\cite{Avigad2019, inp:BoehneKnobelsdorfKreitz16a, Lurch2013b}
    \item Proof assistants help students learn how to structure proofs~\cite{inp:BoehneKnobelsdorfKreitz16a, coqToText, Burke2006, Lurch2013a, Lurch2013b, hendriks2010teaching, Kiehn2017, KnobKreitz2017, SPA}
    \item Students are helped by instant feedback on their proofs~\cite{Avigad2019, inp:BoehneKnobelsdorfKreitz16a, Burke2006, Lurch2013a, Lurch2013b, Lurch2017, SeCaV, KnobKreitz2017, LSDproofsSemantic}
    \item Proof competence gained using a proof assistant transfers to pen-and-paper~\cite{Avigad2019, Lurch2017, LSDproofsSemantic, AProS2007}
    \item Proof assistants help students fix their proof errors as early as possible~\cite{inp:BoehneKnobelsdorfKreitz16a, Lurch2013b}
    \item Proof assistants make it easier for students to get started on a proof~\cite{inp:BoehneKnobelsdorfKreitz16a, SeCaV}
    \item Students consult formal definitions to gain understanding~\cite{Jacobsen2022, Villadsen2015}
    \item Students experiment with formal definitions to gain understanding~\cite{Jacobsen2022, LSDproofsSemantic}
    \item Proof assistants help students experiment with proof ideas~\cite{inp:BoehneKnobelsdorfKreitz16a}
    \item Correcting assignments checked by proof assistants is easier~\cite{Burke2006}
\end{itemize}
Some authors have however also observed various difficulties that students encounter when trying to use proof assistants, including:
\begin{itemize}
    \item Difficulties learning proof assistant syntax~\cite{Avigad2019, Lurch2017}
    \item Difficulties understanding proof assistant error messages~\cite{Avigad2019}
    \item Difficulties transferring proof competencies from proof assistants to pen-and-paper~\cite{coqToText, KnobKreitz2017}
    \item Difficulties stating properties formally~\cite{Burke2006}
    \item Technical difficulties in installing and using a proof assistant~\cite{Burke2006, hendriks2010teaching, KnobKreitz2017}
    \item Difficulties understanding the very expressive language of a proof assistant~\cite{hendriks2010teaching, KnobKreitz2017}
    \item Difficulties remembering formal proof rules~\cite{KnobKreitz2017}
\end{itemize}
Some authors also note potential issues of using a proof assistant in education from the course instructor's point of view, including:
\begin{itemize}
    \item The overhead of introducing a proof assistant to students~\cite{Avigad2019, LSDproofsSemantic}
    \item The need to develop specialized proof scripts for each topic~\cite{inp:BoehneKnobelsdorfKreitz16a}
    \item The need to design exercises that cannot be solved by brute force~\cite{coqToText, Lurch2013b, hendriks2010teaching, KnobKreitz2017, Villadsen2019}
    \item The worry that electronic exams using proof assistants may make it easier to cheat~\cite{hendriks2010teaching, LSDproofsSemantic}
    \item That proof assistants have automation that makes too many exercises trivial~\cite{hendriks2010teaching}
\end{itemize}
Finally, several authors note methodological issues in attempting to provide evidence for any of the above-mentioned claims:
\begin{itemize}
    \item No suitable quantitative measures of proof competence exist~\cite{Avigad2019}
    \item It is difficult to collect evidence about how students interact with proof assistants~\cite{coqToText, Jacobsen2022}
    \item It is unclear how to compare the efficacy of approaches based on proof assistants with approaches based on pen-and-paper approaches in a fair way~\cite{Jacobsen2022, LSDproofsSemantic}
\end{itemize}

\subsection{Approaches From Similar Fields}\label{sec:rel-work-inspo}
Wrenn and Krishnamurthi have developed a tool to enable problem comprehension in functional programming by letting students test their understanding by writing property-based tests against a specification before writing code~\cite{Wrenn2019}, and shown that students will use the tool voluntarily~\cite{Wrenn2020}.
In studies using this tool, they found that students still had several types of misunderstandings that the tool or the lectures could have been improved to alleviate~\cite{Wrenn2021} and that course instructors had blind spots in predicting the misunderstandings students had~\cite{Prasad2022}.
Instrumenting the tool with data collection facilities allowed them to inspect student interactions with the tool in a fine-grained manner, and the tool thus helped instructors discover students' actual misunderstandings, which were different from what the instructors had imagined.

Aleven and Koedinger have designed a so-called Cognitive Tutor, which is a specially designed ``proof assistant'' instrumented with features for tracking student behavior and guiding students to explain their work~\cite{Aleven2002}.
They show that the use of this tool with the guidance features enabled improves student learning outcomes in geometry.
The study was enabled by instrumenting the Cognitive Tutor with facilities for collecting fine-grained data about time spent and steps taken in the exercises completed by students using the tool.

\section{Enabling Educational Research on Proof Assistants}\label{sec:what-is-to-be-done}
As mentioned in the previous section, there are few rigorous studies about the influence of proof assistants on the learning of proof competences, but many claims about their efficacy.
In this section we will discuss the importance of performing studies on the impact of proof assistants in education and the obstacles that we need to overcome to enable these studies.
First, however, we discuss approaches to curtailing the observed difficulties students and instructors encounter when using proof assistants in education.

\subsection{Making it Easier to Learn Proof Assistants}
Proof assistants are generally not developed for students, but for expert users, and the learning curve is thus very steep.
For beginning students, using any formal language can be a big challenge, and the rigorous languages of proof assistants can be even more difficult to break into.
Instructors could potentially decrease the challenge by introducing the language one element at a time and developing tools that can guide the students for the specific learning goals of each activity.
One could imagine this approach combined with adaptive learning tools to guide students at their own pace.
A related issue is that proof assistants typically require users to remember not only the contents, but also the formal names of proof rules.
Course developers could curtail this issue by providing an easy way to see the relevant proof rules in each learning activity without having to look them up in a larger list containing many irrelevant rules.
Students can then engage with the content of the course instead of memorizing names.

Another issue is that students find error messages from proof assistants vague and confusing.
This is typically a result of the language of the proof assistant being so expressive, and implementations being so general, that errors refer to concepts that students are not familiar with, e.g.\ type classes or functors.
In many educational contexts, the full expressive power of the proof assistant is not necessary, and instructors could potentially flatten the learning curve by restricting the language to the fragment required by each activity.
This would allow more specific error messages and perhaps even hints about why a specific application of a proof rule is wrong.

Some students have issues installing and using proof assistants from a purely technical perspective.
Some of the issues could be reduced by developing web-based tools, which have the additional benefit of being easy to update such that students are always on the newest version of the tool in case any issues with the instrumentation are found during the course.

Finally, students have issues moving between proof assistants and traditional pen-and-paper proofs and statements.
This occurs both when formalizing properties stated in natural language and when transferring the proof competences developed with a proof assistant to competences in writing pen-and-paper proofs.
We are not convinced that it is possible to solve these issues by technical means, but approaches where the difference between proof assistant and pen-and-paper is gradually minimized seem promising.

\subsection{Making it Easier to Use Proof Assistants in Education}
Choosing to use a proof assistant in education unavoidably introduces the need to balance the time spent learning the actual course content and the time spent learning to use the proof assistant.
The challenge is to design a course that uses the proof assistant as a tool instead of a course which teaches how to use proof assistants \cite{LSDproofsSemantic} (though a course which aims to do this with intention can of course also be valuable).

When designing exercises which are intended to be solved using proof assistants, there are several potential issues: exercises may be too easy to solve by brute force, automation in the proof assistant may make the exercises trivial and using electronic exercises for assignments or exams may make it easier for students to cheat.
Restricting the language of the proof assistant (as described in the previous section) may also facilitate the design of exercises that are difficult to brute force and remove excessive automation.

The problem of designing exercises and supporting developments for each learning activity remains.
We conjecture that this problem is mainly historical, i.e.\ caused by the fact that many courses have pen-and-paper exercises developed through several years.
When designing new learning activities, we conjecture that using a proof assistant may actually expedite the development of good exercises since obscure or easily overlooked corner cases must be handled during the development and are thus not present to confuse students during the actual course.

\subsection{Gathering Evidence}
One of the main issues with implementing proof assistants in education is that there is little evidence of the efficacy of the approach in improving student proof competences.
It is thus unclear whether spending time and resources on designing course material using proof assistants will actually result in students learning more.
A compounding issue is that it is unclear how to measure proof competence across approaches based on proof assistants and approaches based on pen-and-paper in a quantitative way and without favoring one of the approaches.
When considering studies, instructors must thus be careful to specify the learning objectives they are measuring precisely: is the intention of their course to improve proof competency \emph{in general} or for instance specifically in pen-and-paper proofs?

The other major obstacle to conducting studies of proof assistants in education is that collecting objective, quantitative data about student interactions with proof assistants and learning outcomes is difficult.
Most reports thus rely on surveys which ask students to self-report about their experiences or vague measures based on average grades in the course.
One of the main objectives of \ProofBuddy is to enable researchers to conduct studies of the efficacy of approaches to using proof assistants in education by collecting fine-grained data about the way students interact with proof assistants.

\section{\ProofBuddy}\label{sec:proofbuddy}
\begin{figure}
\includegraphics[width=\textwidth]{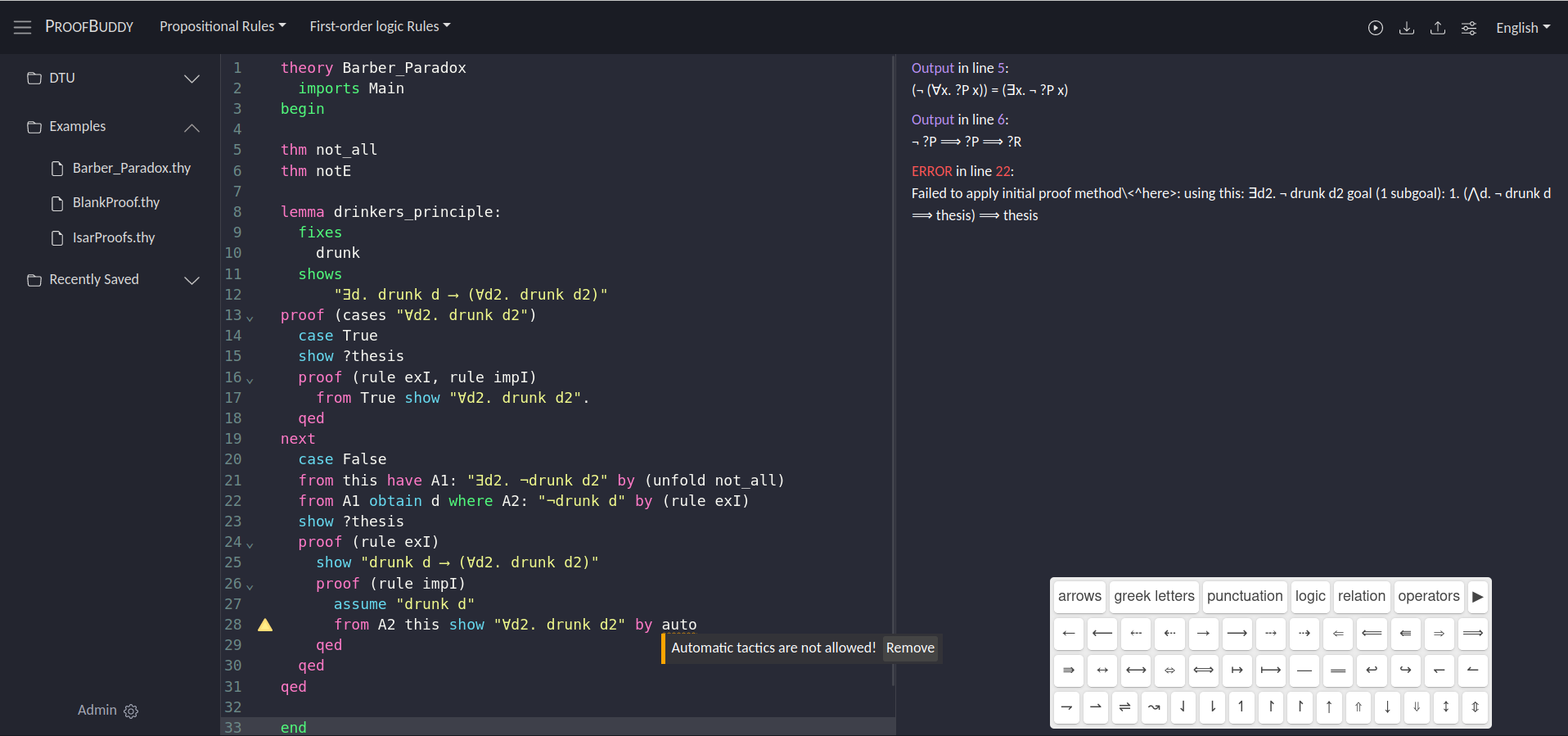} 
\caption{Screenshot of the interface of \ProofBuddy with file browser on the left, editor in the middle and output messages on the right. Note also the linter popup (bottom center) and the symbol keyboard.}
\label{fig:ProofBuddyLinterKeyBoard}
\end{figure}

Inspired by the claimed benefits and drawbacks of using proof assistants for educational purposes described in the previous section, we have developed \ProofBuddy: an instrumented version of the Isabelle proof assistant which is accessible through a web interface.
The tool communicates with a full Isabelle proof assistant running on our server to check proofs and programs and additionally collects data about how users interact with the tool.
It would also have been possible to instrument one of the usual Isabelle editors Isabelle/jEdit or Isabelle/VSCode.
However, that would mean having to update the tool with every new Isabelle version, whereas communicating via the Isabelle server protocol allows us to develop against a more stable target.
Besides the advantage that users do not have to install a special version of Isabelle, a web tool gives us the opportunity to add new features directly on the server without having users reinstall the tool.
Through a management backend on the server it is possible to organize different courses, teacher and students and the collection of data (log-files and the submitted Isabelle theories).
Furthermore a web interface yields more flexibility, and it is conceivable to add other languages like Coq or Lean while preserving a unified interface.
\autoref{fig:ProofBuddyLinterKeyBoard} shows a screenshot of \ProofBuddy.

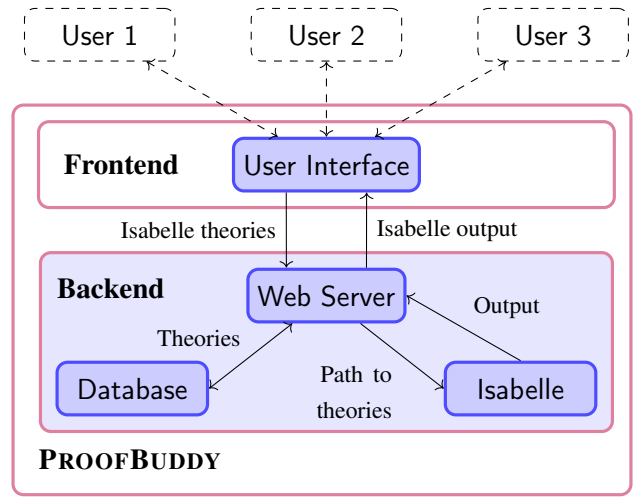
\begin{figure}[t]
\begin{center}
    \begin{tikzpicture}[
        greynode/.style={module, draw=blue!70, fill=blue!20, very thick},
        background-node/.style={module, draw=purple!50, very thick, inner sep = 2mm},
    ]

    \node[module, dashed] (C1) {User 1}; 
    \node[module, dashed, right = of C1] (C2) {User 2}; 
    \node[module, dashed, right = of C2] (C3) {User 3}; 
   
    \node[greynode, below = of C2] (I1) {User Interface};
    \node[greynode, below = of I1] (I3) {Web Server};
    \node[greynode, below right = 7mm of I3] (I5) {Isabelle};
    \node[greynode, below left = 7mm of I3] (I4) {Database};

    \node[above = 7mm of I4, align=left, text width=2cm] (b) {\textbf{Backend}};
    \node[below = 3mm of I4, align=left, text width=2.5cm] (p) {\textbf{\ProofBuddy}};
    \node[left = 1mm of I1, align=left, text width=2cm] (f) {\textbf{Frontend}};
    \node[right = 1mm of I1, align=left, text width=2cm] (f2) {};

    \begin{scope}[on background layer]
        \node[fit = (I3) (I4) (I5), background-node, fill=blue!10] (fit3) {};
        \node[fit = (I1) (f) (f2), background-node] (fit2) {};
        \node[fit = (fit3) (fit2) (I1) (p), background-node] (fit1) {};
    \end{scope}

   \path (I1.south west) -- (I1.south) coordinate[pos=0.575] (I1-mid-sw);
   \path (I1.south east) -- (I1.south) coordinate[pos=0.575] (I1-mid-se);
   \path (I3.north west) -- (I3.north) coordinate[pos=0.5] (I3-mid-nw);
   \path (I3.north east) -- (I3.north) coordinate[pos=0.5] (I3-mid-ne);
   \path (I3.south west) -- (I3.south) coordinate[pos=0.575] (I3-mid-sw);
   \path (I3.south east) -- (I3.south) coordinate[pos=0.575] (I3-mid-se);
    \draw[->] (I1-mid-sw) -- node [left] {\footnotesize Isabelle theories} (I3-mid-nw);
    \draw[->] (I3-mid-ne) -- node [right] {\footnotesize Isabelle output} (I1-mid-se);
    \draw[->] (I3-mid-se) -- node[below left, align=right, text width=1cm] {\footnotesize Path to\\theories} (I5.west);
    \draw[<->] (I3-mid-sw) -- node[above left] {\footnotesize Theories} (I4.east);
    \draw[->] (I5.north) -- node[above right] {\footnotesize Output} (I3.east);
    \draw[dashed, <->] (C1) -- (I1);
    \draw[dashed, <->] (C2) -- (I1);
    \draw[dashed, <->] (C3) -- (I1);

\end{tikzpicture}
\end{center}
\caption{Architecture of \ProofBuddy. Note that each user has their own instance of the frontend running in their browser, while there is only a single backend.}
\label{fig:architecture}
\end{figure}

\ProofBuddy is a web application and hence divided into two parts: the frontend, running in the browser, and the backend, running on the server.
\autoref{fig:architecture} outlines the system architecture of \ProofBuddy.
The design decision to use Isabelle in the backend instead of the frontend came as a result of Isabelle’s development languages, ML and Scala, which are difficult to run in a browser.
Furthermore it is easier to log the communication with the Isabelle server in the backend and save the theories which are sent to be verified.
The frontend and backend of our application communicate via Socket.IO \cite{socketio}, allowing for instantaneous and bi-directional data transfer. 
Socket.IO is an event-driven library based on the WebSocket protocol \cite{fette2011websocket}.
In browsers where WebSockets are not supported, Socket.IO assures a stable connection via HTTP long-polling.

\subsection{Frontend}
The user interface of \ProofBuddy mainly consists of three panes: an editor pane, an output pane and a PDF viewer.
The \emph{editor pane} is the main interactive component.
An Isar proof of the ``drinker's principle'' is shown in the \emph{editor pane} in the middle of \autoref{fig:ProofBuddyLinterKeyBoard}.

\ProofBuddy includes a parser for the language of Isabelle which enables syntax highlighting, autocompletion, code folding, bracket closing, search and replace functionality, etc.
The editor of \ProofBuddy thus functions like a typical integrated development environment (IDE).
Since Isabelle is solely running in the backend, \ProofBuddy does not have the ability of Isabelle/jEdit and Isabelle/VSCode to look up definitions and display type information by clicking on names of e.g.\ functions and theorems.

\begin{figure}
\includegraphics[width=\textwidth]{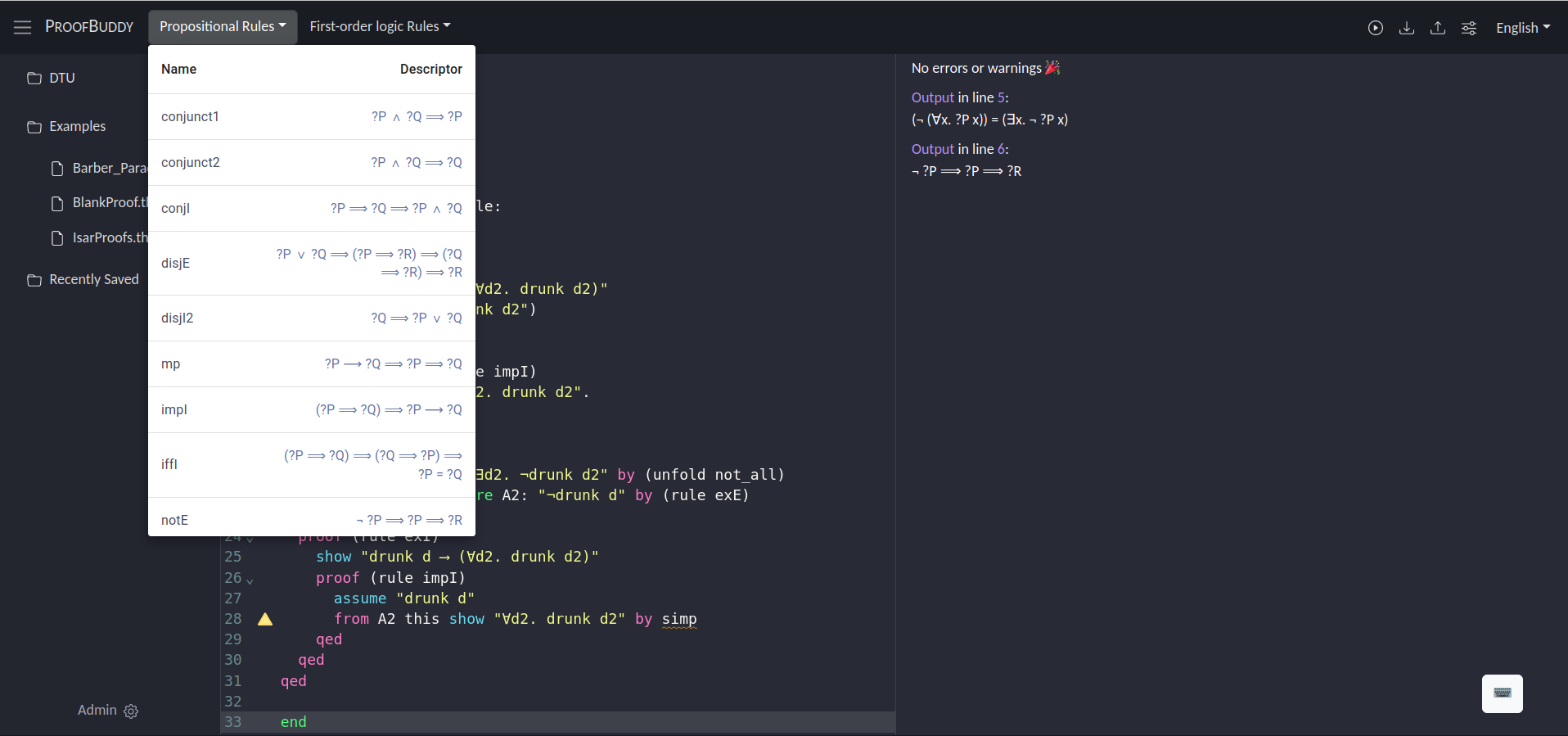} 
\caption{Screenshot of the interface of \ProofBuddy with a reference list of proof rules. Note the yellow warning triangle and yellow underline of the ``simp'' tactic from the linter.}
\label{fig:ProofBuddyRules}
\end{figure}

Instructors can restrict the language, and thus expressivity, of Isabelle on an activity-to-activity basis by adding linters, which disallow certain syntactic constructs.
Linters display errors and warnings instantly in form of popups directly at the point where the failure occurs. 
We have implemented a linter based on regular expressions (as shown in \autoref{fig:ProofBuddyLinterKeyBoard} and \autoref{fig:ProofBuddyRules}) to warn about the usage of the automatic tactics auto, simp, arith and blast. 
At this point of development, the linters do not prevent users from asking Isabelle to check their work even if a linter detects the usage of prohibited syntax for the activity at hand.
We could also restrict syntax by hiding definitions via a prelude, but this would not give users any specific information about their mistake when attempting to use e.g.\ a prohibited tactic, since the error from Isabelle would simply be about attempting to use an undefined tactic.
Students can currently toggle the linter for automatic tactics through a switch on the user interface, but this feature can be removed to force students to only use automation in specific activities.
It is thus possible to introduce features one at a time, ensuring that students can only use the features that they are supposed to learn in a given activity.

The \emph{output pane}, located on the right-hand side, displays the feedback from Isabelle regarding the correctness of the current development.
The closable \emph{PDF viewer} allows the display of tutorials, lecture notes, exercise descriptions, etc.\ within the tool.

The collapsible \emph{sidebar} contains a file browser and the profile of the current user.
Users must log in to access \ProofBuddy.
User profiles are used to store progress and access previously created theories as well as to track interactions of individual users in the collected data.
We did however implement a guest login to allow testing of \ProofBuddy. 
The \emph{toolbar} contains buttons to run the development (i.e.\ send the content of the editor to Isabelle for checking) and to download and upload Isabelle theories.
Additionally, the toolbar contains dropdown menus which list the names and definitions of relevant proof rules on an activity-to-activity basis, as shown in \autoref{fig:ProofBuddyRules}.
Users can insert rule names at the current cursor position in the editor by clicking on a rule in the list.

Users can open the \emph{symbol keyboard} by clicking the keyboard-button in the bottom right corner as seen in \autoref{fig:ProofBuddyRules}.
As shown in \autoref{fig:ProofBuddyLinterKeyBoard}, the keyboard offers an easy way to enter relevant mathematical and logical symbols, which instructors can configure on an activity-to-activity basis.

\subsection{Backend}
The backend of \ProofBuddy is responsible for handling user authentication and data management, but its main functionality is to check the correctness of received developments using Isabelle.

The Isabelle proof assistant usually runs as two processes: a daemon, the Isabelle server, and an interactive application, the Isabelle client, through which theories can be sent to be checked by the server \cite{wenzel2014isabelle}.
The Isabelle server listens on a TCP socket and allows for bi-directional communication with multiple clients following a protocol of structured messages.
The \ProofBuddy backend communicates with Isabelle via the Isabelle client by sending requests to check the developments it receives (checking is incremental as in the usual Isabelle development environments, so only theories with changes are checked).
The messages consist of the session ID and a number of theories.
The Isabelle server acknowledges the request and answers asynchronously whenever it finishes a task.
The message of the Isabelle server includes all the usual error messages which are known from Isabelle/jEdit.
Additionally, the Isabelle server periodically sends progress reports on ongoing tasks.
The \ProofBuddy backend starts the Isabelle client as a child process after ensuring that the Isabelle server is running.
By writing to the input stream and reading the output stream of the Isabelle client, the backend is able to manage Isabelle sessions and check theories using Isabelle while simultaneously logging the interactions.

Isabelle sessions \cite{wenzel2014isabelle} consist of a collection of related Isabelle theories and build upon other Isabelle sessions.
It is necessary to add Isabelle theories to an Isabelle session in order to check them using the Isabelle server.
The current version of \ProofBuddy always uses Isabelle/HOL as the parent session.
Hence, \ProofBuddy makes the entire functionality of Isabelle/HOL available.
For \ProofBuddy (and thus Isabelle) to distinguish theories with the same name that are written by different users, each user has an individual Isabelle session.
Adding theories to an Isabelle session automatically results in Isabelle checking the correctness and completeness of the theories.
Once the Isabelle server has checked a development, the web server sends feedback from Isabelle directly to the frontend using Socket.IO, where it is formatted and displayed to the user.

\subsection{Advantages of \ProofBuddy}
We designed \ProofBuddy in response to the issues described in \autoref{sec:what-is-to-be-done}, and we will here highlight some of the ways in which \ProofBuddy provides opportunities to alleviate these issues.

Due to the design decision of implementing a web interface, students do not need to install anything, and instructors can be certain that all students are using the same version of the tool.

The parser and the linters offer the possibility to restrict the input language used in \ProofBuddy.
The restriction of the input language makes it possible to introduce features one at a time, ensuring that students can only use the features that they are supposed to learn about in a given activity.
The dropdown menus containing reminders of proof rules relevant to the activity at hand could also make it easier for students to engage with the content of the course by eliminating time spent looking up proof rules.
Restricting the input language additionally could make it easier to design exercises that are harder to solve by brute force, e.g.\ by disallowing automation or certain proof rules.
Restricting the input language could also be used to force students to write proofs in a style that resembles pen-and-paper proofs.

By formatting and specializing the messages from Isabelle in the frontend before displaying them, it is possible to give better feedback as described in \autoref{sec:what-is-to-be-done}.
This would make it possible to add more, and also more useful, feedback to the output, depending on the failure or the previous behavior of the user.

Instructors can develop an interactive tutorial which forces students to complete exercises in a certain order by structuring the activities in \ProofBuddy using the built-in file browser and PDF viewer.
This also allows for tailoring exercises to the pace and needs of the individual student.

The major benefit of \ProofBuddy is the ability to monitor the progress and behavior of each student as they learn to program and prove in Isabelle/HOL.
Therefore we are able to collect data to evaluate the behavior of students.
This data consists of the contents of student theories saved whenever the student sends the theory for checking, annotated with time stamps.
Monitoring the individual keystrokes and cursor position could also be possible to get even more detail.
A high frequency of checking a theory can indicate that the student tries something without really thinking about why a proof is not accepted.
If many students struggle with the same exercise, the instructor can improve the exercises or give a better introduction.
When analyzing the failure it is also possible to adapt the next exercises.
Using \ProofBuddy offers two opportunities: (1) educational studies about the impact of proving with proof assistants can be performed with the help of the collected data and (2) learning analytics to improve and adapt the exercises to each student depending on the failures.
In the next section, we will evaluate the extent to which \ProofBuddy is useful for the first of these opportunities.

\section{Evaluation of \ProofBuddy}\label{sec:evaluation}
We evaluate \ProofBuddy both from the perspective of researchers wanting to carry out didactic studies using data collected by the tool and from the perspective of students using the tool.
Since we are not aware of any rigorous studies of how proof assistants impact learning, we are unable to test the adequacy of \ProofBuddy with regards to concrete research questions from the literature, but instead evaluate \ProofBuddy from the perspective of researchers by testing the adequacy of the data collected by \ProofBuddy for answering research questions that we imagine could be part of future didactic studies.
We note that \ProofBuddy can of course only be used to answer questions about the behavior of students when interacting with a proof assistant.

Any comparative study between approaches based on proof assistants and approaches based on pen-and-paper would still need additional work to answer questions about the behavior of students when writing proofs on paper.
Note also that this evaluation is not itself a didactic study, but simply an evaluation of the usefulness of the \ProofBuddy tool for conducting future studies.
The evaluation from the perspective of students concerns the usability of \ProofBuddy, both in general and as compared to the standard editors distributed with Isabelle.

\subsection{Didactic Context}
We carried out the evaluation of \ProofBuddy as part of the existing MSc-level course \emph{Automated Reasoning} at the Technical University of Denmark (DTU) in the spring of 2023.
This is an elective course for advanced Computer Science students, and the main topic is learning the use and theory of Isabelle.
We thus expect students following the course to already be proficient at pen-and-paper proofs, and the main purpose of evaluating \ProofBuddy using a population of students following the course is not to teach the students, but to collect data about how students interact with \ProofBuddy.
\autoref{tab:course-plan} contains a brief overview of the course plan and more information about the course can be found at \url{https://kurser.dtu.dk/course/2022-2023/02256}.
We carried out the evaluation in week 7 of the course, at which point the students were already somewhat familiar with the use of Isabelle.

\begin{table}
    \centering
    \begin{tabular}{r@{ }l|l}
       \multicolumn{2}{c}{Weeks} & Topics \\\hline
       1 to & 2  & Introduction, programming and proving, sequent calculus \\
       3 to & 4  & Logic and proof beyond equality, natural deduction (first-order logic) \\
       5 to & 6  & Isar: a language for structured proofs, natural deduction (higher-order logic) \\
       7 to & 9  & Simple type theory \\
      10 to & 13 & Formalized mathematics and computer science
    \end{tabular}
    \caption{Course plan for the spring 2023 version of \emph{Automated Reasoning} at DTU.}
    \label{tab:course-plan}
\end{table}

\subsection{Research Questions}\label{sec:rqs}
We evaluated the usefulness of \ProofBuddy by attempting to answer the following research questions (RQ), which cover both perspectives of the evaluation:
\begin{description}
\item[RQ1] How many resources does \ProofBuddy need to be usable by many students at once?
\item[RQ2] What functionality do students miss in \ProofBuddy compared to Isabelle?
\item[RQ3] What issues did students encounter when interacting with \ProofBuddy?
\item[RQ4] Is the data collected by \ProofBuddy useful for conducting didactic studies?\\
    Sample research questions in didactic studies include:
    \begin{description}
    \item[SQ1] Which types of mistakes do students make while writing functional programs?
    \item[SQ2] Which types of mistakes do students make while writing proofs?
    \item[SQ3] Which types of feedback are immediately useful to students?
    \item[SQ4] How often do students use the possibility to get feedback from the type checker?
    \item[SQ5] How long do students need for an exercise?
    \end{description}
\end{description}
Research question 4 includes a number of sample questions (SQ) which we imagine that researchers could want to answer when conducting a didactic study.
It is important to note that obtaining answers to these questions would not by itself show anything about the efficacy of proof assistants in education as compared to pen-and-paper.
Instead, answering these questions for multiple groups of students taught using different approaches (e.g.\ various approaches to using proof assistant or with pen-and-paper) could provide comparable measures which could potentially be used to show differences in efficacy.
\ProofBuddy enables such studies by allowing researchers to collect data about student interaction with proof assistants, but other approaches are still needed to collect data about student behavior using pen-and-paper, e.g.\ to conduct a randomized controlled trial to determine the efficacy of using proof assistants as compared to pen-and-paper.

\subsection{Methods}
We conducted the evaluation by asking the participants to solve a number of exercises using \ProofBuddy and then fill out a short questionnaire.
The exercises consisted of two parts: proving formulas directly in the natural deduction system of Isabelle/Pure, and programming and proving the correctness of simple program optimizations for an imperative programming language.
The first and second authors instructed and supervised the students during the evaluation.

\subsubsection{Population}
The population studied in the evaluation consists of the students enrolled in the spring 2023 version of the course \emph{Automated Reasoning}, of which there were 19.
We included only those students who were physically present at the exercise session on March 17 in the evaluation.
We recruited students by asking them to participate before the exercise session.
Participants were not reimbursed financially or otherwise.
Before starting the evaluation, we briefed participants about the purpose of \ProofBuddy and the overall elements of the evaluation, including the research questions and the extent of data collected.
Participants were not briefed about the interface of \ProofBuddy such that we could study the discoverability of the features.
Participants were not debriefed after the evaluation.
12 students opted to participate in the evaluation.

\subsubsection{Ethical Considerations}
We did not collect any personal identifiable information of participants during the evaluation, but did collect all text entered into \ProofBuddy.
We briefed participants about the extent of the data collection before the start of the evaluation.
We considered students to have given informed consent to participate in the evaluation when they had listened to the briefing, read a letter describing their rights as participants, and started using \ProofBuddy.
Students in the course were free to choose not to take part in the evaluation and instead solve the exercises using their usual means instead of \ProofBuddy.
The exercises used for the evaluation were not used for grading students.

\subsubsection{Threats to validity}
Since the population consists of students who have voluntarily selected to follow the course \emph{Automated Reasoning}, we may experience some selection bias.
More importantly, the students following the course already had some experience with Isabelle, and might not experience the same issues as complete beginners.
Our preliminary usability study is thus not generalizable to beginners.
Additional selection bias may be introduced since there may be a correlation between students who actively follow the course (and so are present during the evaluation) and those who do not.
The use of a questionnaire where we ask participants to report their own opinions on the importance of missing features and technical issues with \ProofBuddy may introduce self-report bias.
Manually categorizing participant questions asked during the evaluation may introduce researcher bias.

\subsubsection{Analyses}
To answer RQ1, we monitored the performance of the \ProofBuddy tool and the resource usage (in terms of CPU and memory usage) of the server hosting the tool while performing the evaluation.

To answer RQ2 and RQ3 we asked participants to fill out a questionnaire about their experiences and any issues they may have encountered while using the \ProofBuddy tool.
We additionally recorded the questions participants asked the instructors while using \ProofBuddy.
The questionnaire contained the following questions (questions 1--3 concerning RQ2 and questions 4--5 concerning RQ3):
\begin{enumerate}
\item Do you usually use Isabelle/jEdit or Isabelle/VSCode?
\item Did you miss any features of your usual editor while using \ProofBuddy?
\item If yes, how important do you think each of those features are on a scale from 1 to 5 (with 1 being not important and 5 being very important)?
\item Did you encounter any technical issues while using \ProofBuddy?
\item If yes, how much did each of these issues affect your work on a scale from 1 to 5 (with 1 being not at all and 5 being very much)?
\end{enumerate}

To answer RQ4, we used the data collected by \ProofBuddy during the evaluation to set up mock analyses for the sample questions in \autoref{sec:rqs}.
By setting up analyses for each of the sample questions we determined whether the data collected by \ProofBuddy was adequate to perform analyses in didactic studies.
The data collected by \ProofBuddy includes error messages and warnings from the type checker and the proof checker of Isabelle.
The syntactic, type level and tactic level mistakes students make while writing functional programs and proofs can be categorized using this data.
The data also includes the actual programs and proofs, and can thus also be used to categorize semantic mistakes, i.e.\ programs that type check but do not have correct behavior or proofs that prove a different theorem than what was intended.
SQ1 and SQ2 could be answered by categorizing the mistakes students make and ranking the categories by frequency.
The data collected by \ProofBuddy contains timestamps and includes both the actual theory and any messages from Isabelle.
This data can be used to trace the evolution of the theory over time, noting the messages (e.g.\ errors and warnings) \ProofBuddy has given the user between each step, and thus determining which messages lead to theories with fewer mistakes.
SQ3 could be answered by categorizing the mistakes and messages, then ordering them chronologically and measuring the association between various messages and the disappearance of mistakes in the proofs.
This is of course a simplified view of the causality between messages and mistakes, and studying the actual proof scripts in more detail could enable more refined analyses.
The data collected by \ProofBuddy consists of records of each instance of a student asking the tool for feedback.
SQ4 could thus be answered by counting how many records exist in the dataset for each student.
If we assume that students progress immediately from one exercise to the next, SQ5 could be answered by estimating the time spent on each exercise.

\subsection{Observation Protocol}
Nearly all students started the exercise session using \ProofBuddy.
One student had problems logging in, but trying a new account and password resolved the issue.
At first all students got familiar with the tool and asked questions about the interface.
Most of the students only dealt with the logic exercises, but some started with the program optimization exercise.
During the exercise session we collected the questions the students asked the instructors because there was nearly no interaction between the students.
We sorted the observed questions into the following categories:
\begin{enumerate}
\item \textbf{Creating proofs:}
	Problems during the process of proving, including recognizing and understanding assumptions and subgoals as well as how to start a proof.
    Questions about whether a proof is complete and correct also belong to this category;
\item \textbf{Mathematical inscriptions:}
	Problems with mathematical inscriptions, like how to write a proof step or argument in formal language;
\item \textbf{\ProofBuddy usability:}
	Problems with the web interface of \ProofBuddy, like menu functions and understanding the meaning of displayed elements;
\item \textbf{Working with Isabelle:}
	Problems with Isabelle, like choosing the correct rule, variable or structure and keywords of the Isar language, including typing errors;
\item \textbf{Logic:}
	Problems with aspects of logic, like syntax and semantics of definitions or the usage of quantifiers and other connectives;
\item \textbf{Functions:}
	Problems with aspects of functional programming, like syntax and semantics of functions or the concept of functions.
\end{enumerate}

Next, we summarize the kinds of questions observed in each category:
\begin{enumerate}
\item \textbf{Creating proofs:}
There was only one question related to induction.
\item \textbf{Mathematical inscriptions:}
The students had no questions in this category.
\item \textbf{\ProofBuddy Usability:}
There were 17 question in this category, concerning: whether there is a difference in the syntax of Isabelle and \ProofBuddy, how to use the special symbol keyboard, how to load and check theories (five questions).
\item \textbf{Working with Isabelle:}
There were 19 questions in this category, concerning: syntax of Isabelle, the usage of strings in Isabelle and the structure of induction in Isabelle.
\item \textbf{Logic:}
There were 21 questions in this category, concerning: scope of quantifiers, the freshness of variables when eliminating an existential quantifier (using the obtain keyword) and contradiction.
\item \textbf{Functions:}
There were two questions in this category, concerning: conceptional understanding of functions, their semantics and programming. 
\end{enumerate}

In the first hour, most questions concerned the usage of \ProofBuddy and the checking of theories.
We observed that \ProofBuddy needed a long time to answer the checking requests and therefore the students had to wait for feedback.
For that reason students started using their own installations of Isabelle to check the proofs, but kept writing their proofs using \ProofBuddy.
The students seemed to like interacting with \ProofBuddy, and several noted especially the possibility to choose a rule from the drop-down menu, without searching for the right rule among many irrelevant rules.

After the first half hour there were many questions about the syntax of Isabelle and logic.
After one and a half hours the students worked on their own and asked only a few further questions.
The first students left after two hours, and 6 students were still working after two and a half hours when the evaluation ended.
All participants except one filled out the questionnaire before leaving.

\subsection{Data Analysis}
\begin{figure}
 \begin{minipage}[b]{.48\linewidth}
\includegraphics[width=\textwidth]{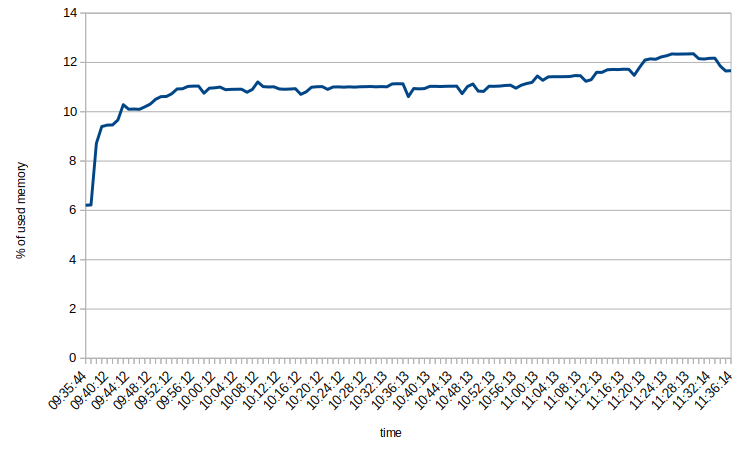} 
\caption{Utilization of the server memory when running \ProofBuddy.}
\label{fig:workload-mem}
\end{minipage}
\begin{minipage}[b]{.48\linewidth}
 \includegraphics[width=\textwidth]{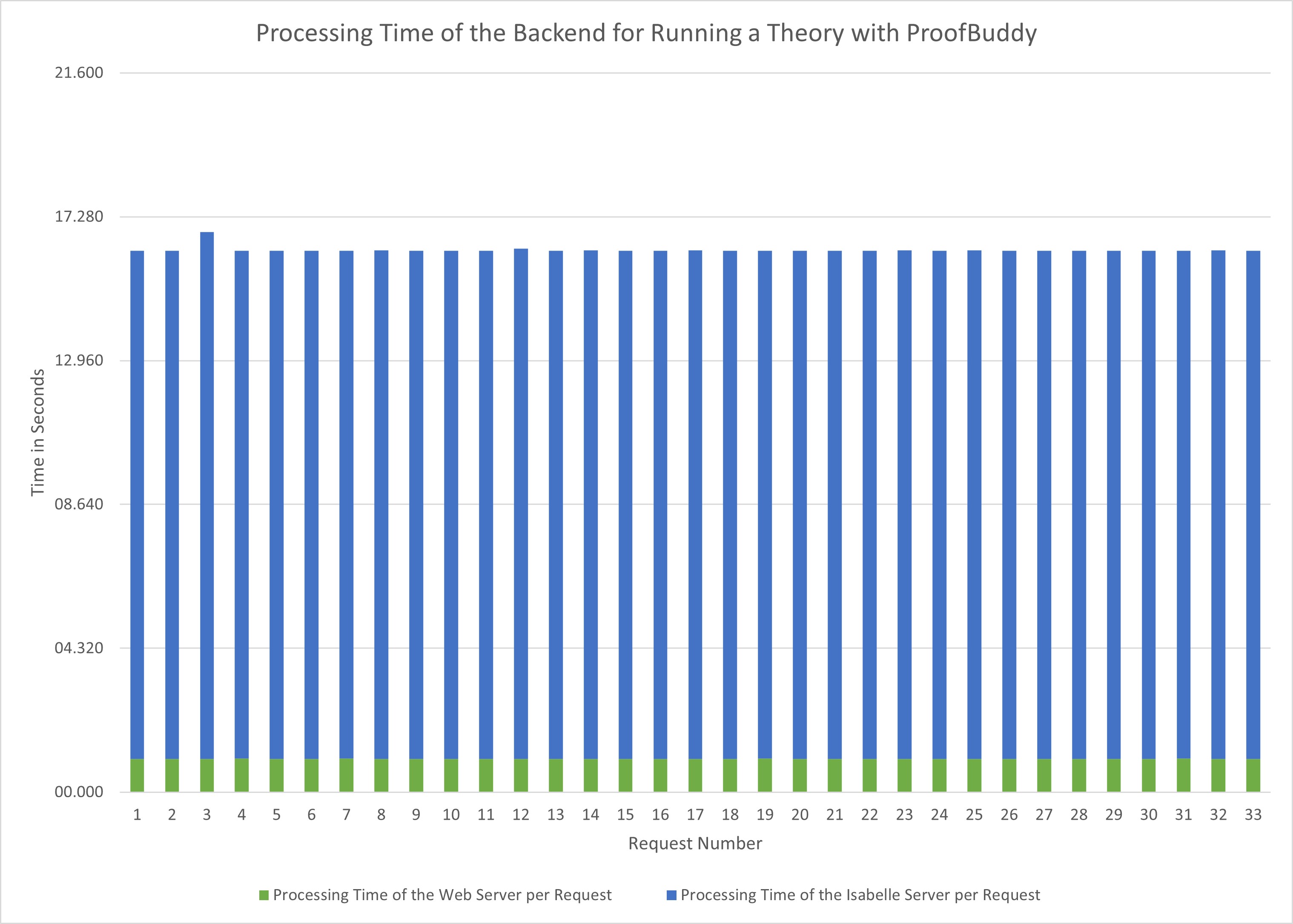} 
\caption{Vertical diagram displaying the processing time of the backend for checking theories with \ProofBuddy.}
\label{fig:backend-processing-times}   
\end{minipage}
\end{figure}

We analyzed the logged communication between frontend, backend and Isabelle Server (see \autoref{fig:backend-processing-times}).
This reveals that the average processing time of a request in the backend takes 16 seconds, where the handling of the request by the web server, i.e.\ writing the development to a theory file and attaching dependencies to the Isabelle request, takes only one second.
The Isabelle Server used the remaining 15 seconds to check the development and compute the feedback.
We did not notice any spikes in the processing time.
Furthermore we analyzed the memory and CPU utilization of the server. 
\autoref{fig:workload-mem} shows the percentage of memory used during the evaluation.
There is an increase in memory usage when starting \ProofBuddy and another one when the students begin to log in.
The measurement of the CPU utilization always resulted in the same value (0.64\% at the user level).

The questionnaire included two different validated standard questionnaires and questions comparing Isabelle and \ProofBuddy.
11 of 12 participants completed the survey and the answers support the impression we got by observing the exercise session.

We used the validated short version questionnaire System Usability Scale (SUS) \cite{brooke1996sus} where \ProofBuddy obtains an average score of 65/100.
The students find \ProofBuddy easy to use but do not see themselves using the tool frequently.

The second questionnaire used is the short version of the User Experience Questionnaire (UEQ) \cite{ueq-s2017}.
In this questionnaire, values between $-0.8$ and $0.8$ represent a neutral evaluation of the corresponding scale, values $>0.8$ represent a positive evaluation and values $<-0.8$ represent a negative evaluation.
The range of the scales is between -2 (horribly bad) and +2 (extremely good).
The pragmatic quality has a value of 0.775 where the efficiency has a bad score and the hedonic quality has a value of 0.727.

The rest of the questionnaire concerned the comparison of \ProofBuddy and Isabelle.
Most of the students normally use Isabelle/jEdit and not Isabelle/VSCode. 
In comparison with Isabelle, the students miss the actual proof state and the immediate feedback and criticize the long waiting time and the error detection.
There was a problem switching between the different activities, which overrides the editor content without saving the previous content anywhere.
Sometimes the cursor jumps to the beginning of the editor.
Often the students do not find the symbol keyboard and sometimes the keyboard overlaps with the editor or output panes.
It sometimes happens that \ProofBuddy does not detect all errors in a theory, even when copying the theory into Isabelle resulted in errors.
Some students used \ProofBuddy to write proofs but checked them in Isabelle/jEdit.
Some students mentioned that \ProofBuddy never gave them any feedback and instead seemed to become stuck when asked to check a theory.

\subsection{Discussion}
As we observed, students like the web interface of \ProofBuddy, especially the dropdown menu.
They find it intuitive to use the interface but there are some issues.

The main issue observed with the usability of \ProofBuddy was the slow checking of theories.
The bad score in efficiency in the UEQ follows from the long waiting time which was the main usability problem. 
Therefore students state in the SUS that they will not use \ProofBuddy frequently.
Additionally, some students reported that \ProofBuddy did not report errors in their theories, indicating another issue with the communication between \ProofBuddy and Isabelle.
We note also that the measurement of CPU utilization always having the same value might have some connection to this issue.

Since the evaluation, the issues concerning the response time have been fixed. Firstly, there was an error in processing and combining chunks of large responses from the Isabelle server, in which case \ProofBuddy appeared to check the theory indefinitely.
Secondly, we specified the headless session option for the delay to consolidate the status of command evaluation (``headless\_consolidate\_delay'').
The ``headless\_consolidate\_delay'', which is by default set to 15 seconds, seems to correlate to the response time of 15 seconds by the Isabelle server.
By reducing the ``headless\_consolidate\_delay'' to 0.5 seconds, we observed a much shorter response time from the Isabelle server and therefore from \ProofBuddy.
During the evaluation, users were only able to access recently saved theories through the file browser.
Reloading the page thus resulted in losing any progress made, since the users were unable to access the saved theories on the server.
Users are now able to access saved theories through the file browser.

Despite the usability issues of \ProofBuddy, the data collection features worked as expected.
All of the data required to answer the sample question as described above was thus present in the \ProofBuddy database.
Unfortunately, the usability issues meant that many students stopped using \ProofBuddy and returned to using their own Isabelle instead.
In a real study this would of course not be acceptable, and so our first priority is remedying the observed usability issues.

\section{Future Work}\label{sec:future-work}
We plan to improve \ProofBuddy.
One approach for optimizing the response time is pre-assessing the Isabelle theories in the frontend.
The frontend is already equipped with a parser for the Isabelle language.
This opens the possibility to assess the theories for syntactical errors, such as missing brackets, or failures in the Isar proof syntax where special keywords are missing, prior to sending them to the backend. 
Only sending syntactically correct theories would reduce the server workload, which could improve the scalability of the \ProofBuddy backend.

We additionally plan to extend the course management functionality by adapting the linter, the drop-down menu and the symbol keyboard to more exercises to support the declared learning objectives of the associated learning activities.
Furthermore, we need to improve the file management, such that there is a database which stores courses and teacher profiles and for each student their own files within their profile.
\ProofBuddy already saves the Isabelle theory versions on the server explicitly, if the user presses the upload button, or implicitly, prior to the checking by Isabelle.

Another step will be to analyze the collected data automatically in order to answer the research questions: (1) what kind of feedback do students need during the proof writing process and (2) which parts of the proof are causing the students difficulties.
By answering the first research question, the feedback of Isabelle can be extended to give more precise feedback depending on the kind of mistake.
The problems discovered by answering the second research question could be solved by a hint to solve an exercise (adaptive learning) or by explaining a concept in the lecture one more time.

We have developed a new BSc-level course at TUB, where we focus on teaching how to create and properly write down proofs.
The course is planned for summer 2023.
Referring to our research questions above, the concept of the course is to enable substantially more feedback for students.
Thus, we use \ProofBuddy in addition to the help of the instructor and fellow students to provide the feedback much more directly (in real time) and also more individually.
The idea is to start with propositional logic and first-order logic, because knowledge about these is a good foundation for structuring proofs \cite{SeldenSelden2009}.
Afterwards, we introduce inductive data structures and prove properties about them.
We want to avoid that students just learn to use the tool without actually understanding the proofs that they develop with it.
Therefore, the concept of the course aims to strengthen the mutual transformation between formal proofs---as developed with \ProofBuddy---and traditional pen-and-paper proofs, in both directions. 
Confronted with these transformations, students are supposed to also learn that there are different degrees of formality to prove propositions.
This course offers the opportunity to evolve the kind of feedback a proof assistant should give to learners that it assists during the learning process.
We plan to iteratively change the feedback during several semesters and analyze if the students have fewer problems with the exercises.
Furthermore we will test new features in a Bachelor course dealing with graph structure theory at TUB in summer 2023, specifically for exercises about tree width.
The students in this course have not used a proof assistant before.
The focus of this study lies on the time students need to step into a proof assistant and how exercises have to be prepared such that students can manage the formal language.

There will also be an introductory course in theorem proving with Isabelle at TUB in summer 2023 where we start with \ProofBuddy to introduce propositional logic, first-order logic and sets. The exercises will be solved in Isabelle/jEdit with the opportunity to use \ProofBuddy instead.

We would also like to carry out a study about the impact of a proof assistant for learning proofs.
This could be measured in an experiment where one group is taught with \ProofBuddy and a control group only learns the topic with pen-and-paper proofs.
But for this kind of study one has to measure the learning of proof competences in a fair way, which can be difficult as detailed above.

\section{Conclusion}\label{sec:conclusion}
We introduce \ProofBuddy, a web-based tool for guiding and monitoring the use of Isabelle/HOL by students.
Proof assistants like Isabelle allow students to write proofs about functional programs, and many concepts in, e.g.\ logic can naturally be encoded as functional programs.

Unfortunately, not much evidence has yet been collected about the efficacy of using proof assistants in education.
We have identified that one of the main issues in doing so seems to be a lack of tools for studying the interactions of students with proof assistants.
Additionally, instructors have reasonable worries about teaching with very expressive languages, which may be difficult to learn and understand.
\ProofBuddy provides the opportunity to restrict the expressivity of Isabelle depending on the exercise and can therefore be used as a learning and teaching tool.
Furthermore \ProofBuddy allows researchers to collect data about student interactions for conducting studies.
We log the data of the communication between frontend, backend and the Isabelle server.
The collected data allows us to improve the error messages of Isabelle and analyze at which parts in a proof the students fail.  
Hence, we can offer hints and additional feedback.
Such an approach allows instructors to gain insight into how students try to write and prove properties about inductive definitions and use pattern matching; it also enables us to carry out didactic research on student behavior.
We expect that instructors could use the results of such studies to design guided exercises which support the learning progression of individual students.

\bibliographystyle{eptcs}
\bibliography{references}

\end{document}